# Superconductivity in In-doped AgPbBiTe$_3$ compounds synthesized by high-pressure synthesis

Takahiro Sawahara[a], Hiroto Arima[a], Takayoshi Katase [b], Aichi Yamashita [a], Ryuji Higashinaka[a], Yoshikazu Mizuguchi[a]*

[a] Department of Physics, Tokyo Metropolitan University, 1-1, Minami–Osawa, Hachioji 192-0397, Japan.
[b] Laboratory for Materials and Structures, Institute of Innovative Research, Tokyo Institute of Technology, 4259 Nagatsuta, Midori, Yokohama 226-8503, Japan.
(*Corresponding author: mizugu@tmu.ac.jp)

**Abstract**

NaCl-type metal tellurides ($M$Te) have been widely studied due to unique physical properties. We investigated the In-doping effects on structural and physical properties of Na-Cl type (AgPbBi)$_{(1-x)/3}$In$_x$Te and the superconducting properties of the In-doped samples. Polycrystalline samples with $x$ = 0–0.5 were synthesized by utilizing high-pressure synthesis. For $x$ = 0.2–0.5, superconductivity was observed in magnetization measurements, where the highest transition temperature ($T_c$) was 2.8 K for $x$ = 0.4. We measured specific heat for $x$ = 0.4 and confirmed the bulk nature of the superconductivity. The evolution of the Seebeck coefficient and lattice constant by In doping suggests that In valence state is In$^{+3}$, and the In doping generates electron carriers in the (AgPbBi)$_{(1-x)/3}$In$_x$Te system.

**Keywords**

New superconductor
Metal telluride
High-pressure synthesis
Carrier doping

## 1. Introduction

Metal tellurides (*M*Te) with a NaCl-type structure, such as SnTe, PbTe, and InTe, have been extensively studied because of their unique physical properties as a topological insulator [1-4], thermoelectric materials [5-9], and superconductors [10-24]. In SnTe, a topological crystalline insulator, superconductivity is induced by career doping by partial substitutions of In or Ag for the Sn site [12-17]. Without those doping, SnTe shows low-temperature superconductivity below 0.3 K due to Sn defects ($Sn_{1-x}Te$) [12-13], but the In and Ag doping increases the superconducting transition temperature ($T_c$) to 4.2 and 2.4 K, respectively [16-17]. In particular, In-doped SnTe has been studied as a candidate of a topological superconductor [14-16]. Therefore, further development of new NaCl-type telluride superconductors has been desired to expand the material variation. In this work, we used alloying technique to design new superconductors. In a previous report, ($Ag^+_{0.5}Bi^{3+}_{0.5}$) was used to replace the Sn site in SnTe, which preserves the crystal-structure type and the total valence states [18]; ($Ag_{0.5}Bi_{0.5}$) could be considered as a virtual divalent atom. By substituting ($Ag^+_{0.5}Bi^{3+}_{0.5}$) for Sn, lattice constant ($a$) can be systematically controlled, but no carriers were doped. For SnTe-based materials, band calculation suggested that the band gap changes with lattice constant, and a band inversion transition occurs at around $a$ = 6.35 Å. In terms of thermoelectric materials, substituting ($Ag^+_{0.5}Bi^{3+}_{0.5}$) for Sn lowers thermal conductivity then a high thermoelectric figure $ZT$ is achieved [25]. Furthermore, equimolar $AgSnBiTe_3$ exhibits superconductivity by In doping [18]. As another *M*Te with a NaCl-type structure, PbTe has been widely studied as a thermoelectric material [5-7], and PbTe shows superconductivity below 5 K after In doping for the Pb site [19]. From the analogy between SnTe and PbTe, ($Ag^+_{0.5}Bi^{3+}_{0.5}$) could be partially substituted at the Pb site like in SnTe, and $AgPbBiTe_3$ was studied as a thermoelectric material [8]. However, heavy carrier-doping effects with an aim of the emergence of superconductivity have not been studied in $AgPbBiTe_3$.

$AgPbBiTe_3$ does not exhibit superconductivity at ambient pressure. By applying pressure ($P$), superconductivity is induced at $P$ = 2.6 GPa, and the highest $T_c$ was 6.5 K at $P$ ~ 25 GPa [20]. Two-step structural phase transitions occur in $AgPbBiTe_3$ at $P$ ~ 10 GPa (from NaCl-type to *Pnma*) and at $P$ = 14.28 GPa (from *Pnma* to CsCl-type) [20]. In the CsCl-type phase, band gap is suppressed, and $AgPbBiTe_3$ becomes metallic. Therefore, to induce superconductivity at ambient pressure, heavy carrier doping is needed. Here, we studied In-doping effects on structural and physical properties in $AgPbBiTe_3$ and investigated the superconductivity properties of In-doped compositions.

## 2. Experimental details

Polycrystalline sample of $(AgPbBi)_{(1-x)/3}In_xTe$ with $x$ = 0 was prepared by a melting method in an evacuated quartz tube under ambient pressure. Powder of Ag (99.9%up) and grains of Pb (99.9%), Bi (99.9999%) and Te (99.999%) were mixed with normal compositions and melted in an evacuated quartz tube; the quartz tube was heated at 800°C for 15 h. The obtained sample was powdered, pelleted, sealed in an evacuated quartz tube, and heated at 400°C for 15 h. For $x$ = 0.1, 0.2, 0.3, 0.4, and 0.5, high-pressure annealing of respective precursor powders was used to obtain In-doped samples. InTe under ambient pressure is orthorhombic and not cubic NaCl-type; with the effects of external pressures, NaCl-type structure is achieved [19]. The precursor powders were synthesized by a melting method. In addition to the above materials, grains of In (99.99%) were used. The samples were heated at 800°C for 20 h, and the obtained sample was powdered and pelleted into a diameter of 5 mm. The pellet was set in a high-pressure cell consisting of a BN sample capsule, carbon heater capsule, electrodes, and pyrophyllite cubic cell. The high-pressure annealing was performed using a cubic-anvil-type 180-ton press (CT factory) under $P$ = 3 GPa at 500°C for 30 minutes. The high-pressure annealing condition was optimized after testing various temperature and pressure conditions.

To examine the phase purity and crystal structure of the obtained (AgPbBi)$_{(1-x)/3}$In$_x$Te samples, powder X-ray diffraction (XRD) was performed by the $\theta$-2$\theta$ method with a Cu-K$\alpha$ radiation using a MiniFlex600 diffractometer (RIGAKU) equipped with a high-resolution detector D/tex-Ultra. The obtained XRD patterns were refined by the Rietveld method using Jana2020 software [26] to estimate the crystal structural parameters. The schematic image of the crystal structure was created using VESTA software [27]. The actual chemical composition of the samples was analyzed by an energy-dispersive X-ray spectroscopy (EDX) with a SwiftED analyzer (Oxford) equipped on a scanning electron microscope TM-3030 (Hitachi Hightech).

The temperature dependence of magnetic susceptibility was measured using a superconducting quantum interference device (SQUID) on a magnetic property measurement system (MPMS3, Quantum Design) at zero-field-cooling (ZFC) and field-cooling (FC) conditions with a magnetic field of $B$ = 10 Oe. The temperature dependence of electrical resistivity was measured by a four-probe method with an applied DC current of 5 mA on a Physical Property Measurement System (PPMS, Quantum Design) under magnetic fields. The temperature dependence of specific heat was measured by a relaxation method on the PPMS under 0 and 5 T using a $^3$He probe system (Quantum Design). Seebeck coefficient ($S$) at room temperature was measured by a four-probe method with a ZEM-3 (Advance Riko) instrument for $x$ = 0 and on a prober system for $x$ = 0.1–0.5. The thermo-electromotive force ($\Delta V$) and the temperature difference ($\Delta T$) were measured simultaneously, and $S$ was estimated from the liner slope of $\Delta V/\Delta T$.

## 3. Result and discussion

The actual compositions of the obtained samples estimated by EDX are listed in **Table 1**. The estimated compositions are reasonably consistent with the nominal values. Hence, we use nominal $x$ to discuss results in this paper.

**Figure 1(a)** shows the powder XRD patterns for (AgPbBi)$_{(1-x)/3}$In$_x$Te ($x$ = 0–0.5). For all samples, the XRD patterns coincided with that of a NaCl-type structure (space group: $Fm$-3$m$, #225, O$_h^5$), as shown in **Fig. 1(b)**. For $x$ = 0–0.3, single-phase samples were obtained, although the XRD peaks of $x$ = 0.3 are broader than of $x$ = 0–0.2. For $x$ = 0.4, tiny peaks are observed at 2$\theta$ slightly lower than the major peaks, which indicates the presence of a NaCl-type impurity phase with a larger lattice constant. As $x$ increased to 0.5, the impurity peaks became clear. Similar trend was reported in (AgSnPbBi)$_{(1-x)/4}$In$_x$Te [21], and the appearance of impurity is caused by solubility limit of In in the $M$Te structure. This trend is understood by the fact that InTe does not have a NaCl-type structure by orthorhombic structure at ambient pressure. Only when the effects of external pressure were applied, InTe becomes a NaCl-type compound. In the current system, the solubility limit of In in (AgPbBi)$_{(1-x)/3}$In$_x$Te is between $x$ = 0.4 and 0.5. With increasing $x$, the major peaks shift to higher angles, which indicates that the lattice constant decreases with increasing In concentration. **Figure 1(c)** shows the nominal $x$ dependence of the lattice constant $a$, which was estimated by Rietveld refinements. Lattice constant decreases with increasing $x$; a similar trend was observed in other NaCl-type $M$Te [16,18,19,21]. For $x$ = 0.3, the lattice constant deviates from the trend of other compositions. This would be caused by the broad XRD peaks for $x = 0.3$, which suggests inhomogeneity of compositions in the $x = 0.3$ sample.

In solid solution systems, lattice constant could be calculated by Vegard's law. For $(AgPbBi)_{(1-x)/3}In_xTe$, $a$ is calculated in following formula:

$$a = a_1(1-x) + a_2x,$$

where $a_1$ and $a_2$ are lattice constant of $AgPbBiTe_3$ and $InTe$, respectively. This function is shown by the dashed line in **Fig. 1(c)**. None of the lattice constants agreed with the Vegard's law. This suggests that the average In valence is smaller than +2 because InTe has evenly mixed valence state with $In^{+1}$ and $In^{+3}$. The ionic radius of $In^{3+}$ is smaller than that of $In^{+1}$. Therefore, in $(AgPbBi)_{(1-x)/3}In_xTe$, the average In valence state is greater than +2.

**Table 1.** Actual composition estimated by EDX, lattice constant by Rietveld refinement, and $T_c$ from magnetic susceptibility ($T_c^{mag}$) for $(AgPbBi)_{(1-x)/3}In_xTe$.

| Nominal $x$ | Actual composition | Lattice constant $a$ (Å) | $T_c^{mag}$ (K) |
|---|---|---|---|
| 0 | $Ag_{0.29}Pb_{0.31}Bi_{0.34}Te$ | 6.2775(2) | - |
| 0.1 | $Ag_{0.34}Pb_{0.28}Bi_{0.30}In_{0.11}Te$ | 6.2426(8) | - |
| 0.2 | $Ag_{0.27}Pb_{0.24}Bi_{0.27}In_{0.21}Te$ | 6.2224(10) | 2.3 |
| 0.3 | $Ag_{0.23}Pb_{0.23}Bi_{0.25}In_{0.32}Te$ | 6.2202(27) | 2.6 |
| 0.4 | $Ag_{0.20}Pb_{0.20}Bi_{0.21}In_{0.41}Te$ | 6.1491(17) | 2.8 |
| 0.5 | $Ag_{0.18}Pb_{0.15}Bi_{0.15}In_{0.55}Te$ | 6.1397(19) | 2.6 |

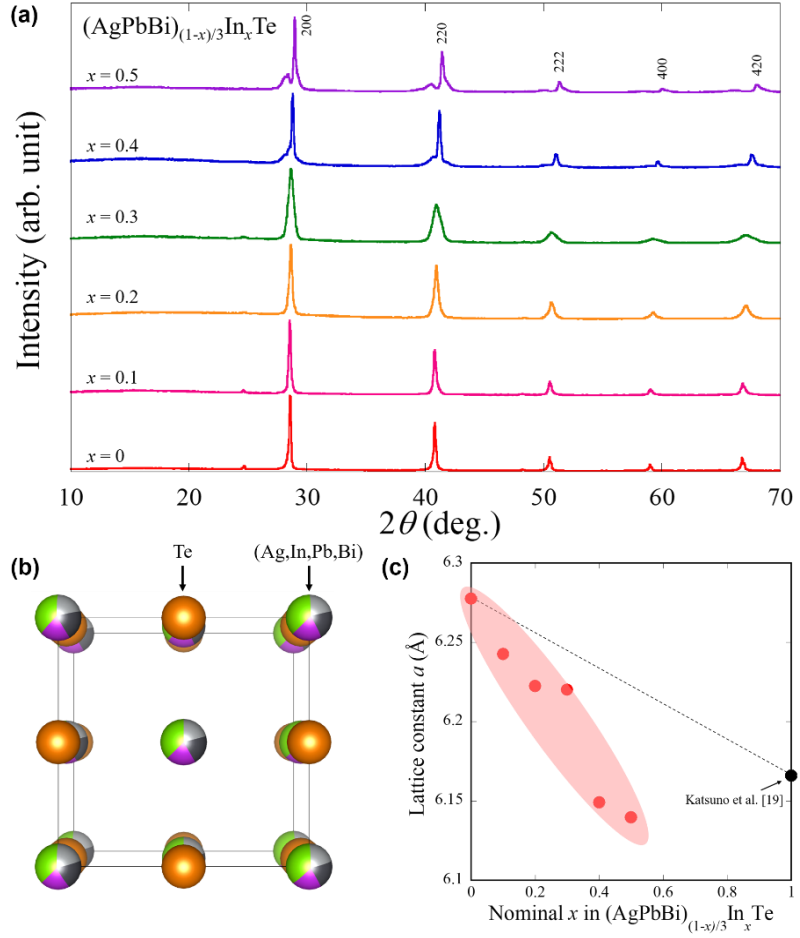

**Fig. 1.** (a) Powder XRD patterns for $(AgPbBi)_{(1-x)/3}In_xTe$. (b) Schematic image of crystal structure for $x = 0.4$. (c) Nominal $x$ dependence of lattice constant.

To investigate the effect of In-doping on superconductivity, the temperature dependence of magnetic susceptibility ($4\pi\chi$) was measured for $x = 0.1$–$0.5$. As reported in Ref. 21, $AgPbBiTe_3$ ($x = 0$) does not exhibit superconductivity above 1.8 K under ambient pressure [21]. For $x = 0.1$, superconductivity was not observed above 1.8 K. For $x = 0.2$–$0.5$, superconducting diamagnetic signals were observed as shown in **Fig. 2(a)**. Each signal showed a large shielding volume fraction, suggesting the emergence of bulk superconductivity. It should be mentioned that the data were not corrected by demagnetization effect. From the susceptibility results, $T_c$ was estimated as a temperature where the shielding volume fraction exceeds 1% ($T_c^{mag}$). **Fig. 2(b)** shows the $x$ dependence of $T_c^{mag}$. As increasing $x$, $T_c^{mag}$ increases and reaches the maximum value of $T_c^{mag} = 2.8$ K at $x = 0.4$. Further increase in $x$ resulted in a decrease in $T_c^{mag}$. This behavior may be due to excess carrier doping and/or the presence of impurity. The In-concentration dependence of $T_c$ in $(AgPbBi)_{(1-x)/3}In_xTe$ is similar to that in $(AgSnBi)_{(1-x)/3}In_xTe$ [18], but the $T_c$ for the present system is higher than $(AgSnBi)_{(1-x)/3}In_xTe$ at all $x$.

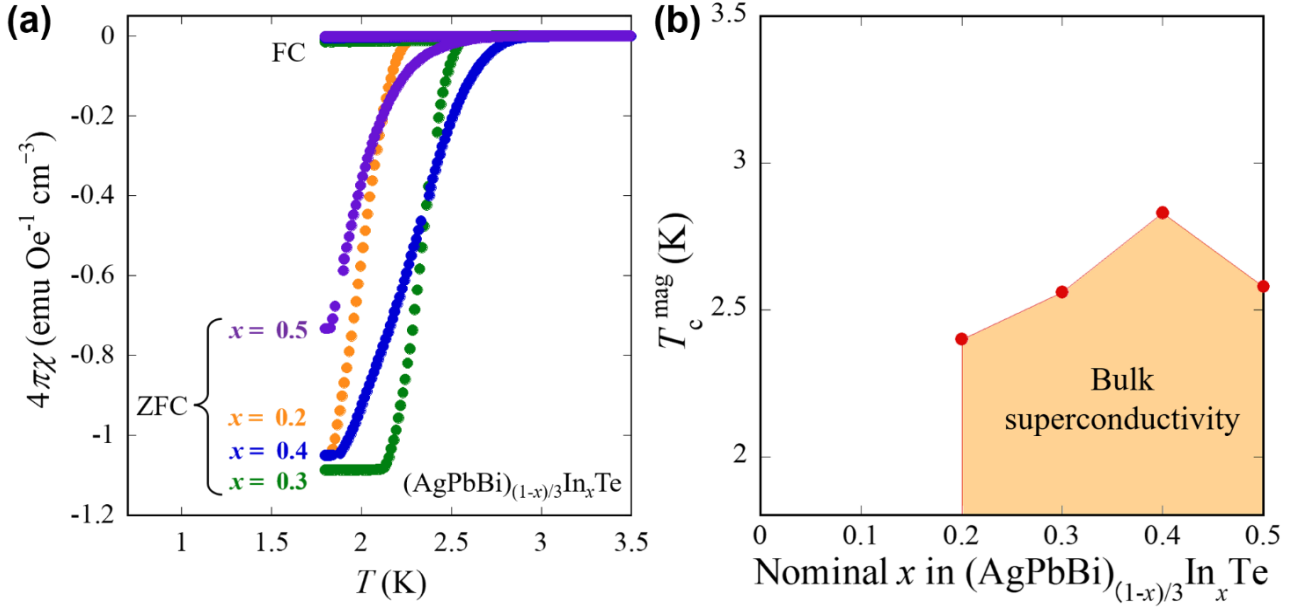

**Fig. 2.** (a) temperature dependence of magnetic susceptibility ($4\pi\chi$) for $x$ = 0.2–0.5. (b) superconductivity phase diagram of $(AgPbBi)_{(1-x)/3}In_xTe$

To investigate the superconductivity properties in the $(AgPbBi)_{(1-x)/3}In_xTe$ system, the temperature dependence of electrical resistivity and specific heat were measured for $x$ = 0.4 with the highest $T_c^{mag}$. **Figure 3(a)** shows the temperature dependence of resistivity of $x$ = 0.4 under magnetic fields up to 0.7 T with an interval of 0.1 T. At $B$ = 0 T, the $T_c^{onset}$ is around 2.7 K, and zero resistivity was observed below $T_c^{zero}$ = 2.6 K. By applying magnetic fields, both $T_c^{zero}$ and $T_c^{onset}$ shift to lower temperatures. $T_c^{zero}$ was not observed at $B$ > 0.3 T, and $T_c^{onset}$ was not observed at $B$ > 0.5 T. To estimate the upper critical field ($B_{c2}$), the estimated $T_c^{onset}$ were plotted in **Fig. 3(b)**. We applied Ginzburg Landau (GL) theory:

$$B_{c2}(T) = B_{c2}(0) \cdot [(1-t^2)/(1+t^2)],$$

where $t = T/T_c$ is the reduced temperature, and $B_{c2}(0)$ is the upper critical field at $T$ = 0 K [28,29]. As a result, $B_{c2}(0)$ was estimated as 1.5 T.

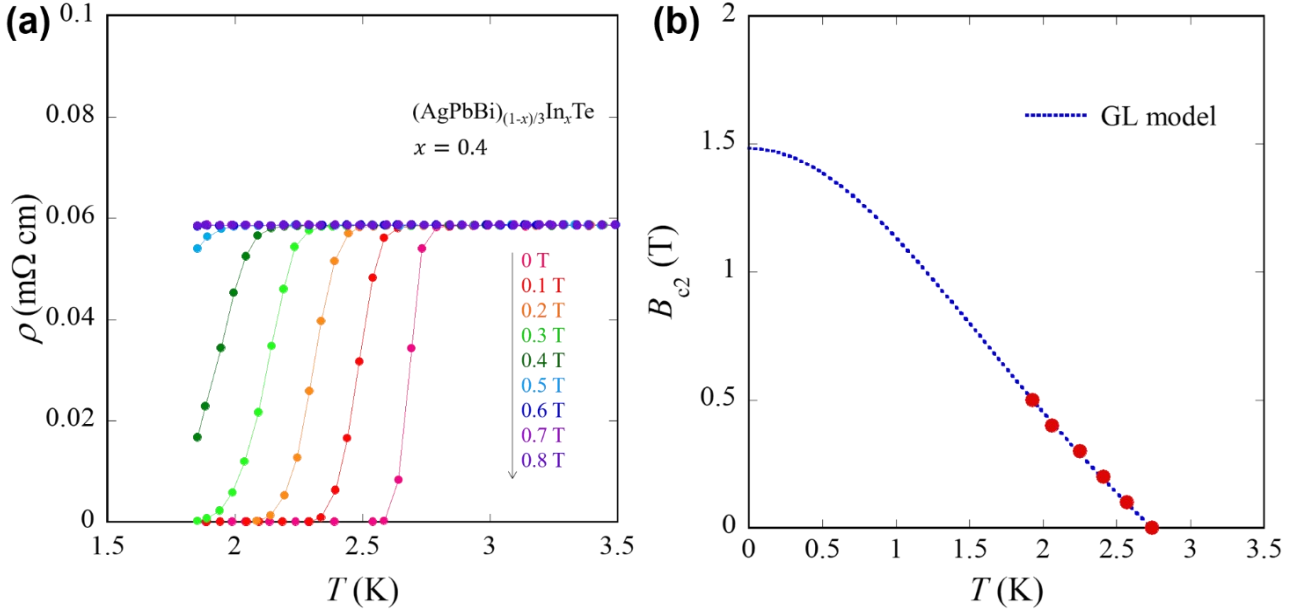

**Fig. 3.** (a) Temperature dependence of electrical resistivity for $x = 0.4$ under magnetic fields of 0–0.8 T. (b) Temperature dependence of $B_{c2}$ for $x = 0.4$. The red plotted points are $T_c^{onset}$. The blue dashed line indicates the GL model fitting.

**Figure 4(a)** shows the $T^2$ dependence of specific heat ($C$) in a form of $C/T$ for $x = 0.4$ under 0 and 5 T. The low-temperature specific heat is described as $C = \gamma T + \beta T^3 + \delta T^5$, where the first, second, and third terms represent contributions from electron, lattice, and anharmonicity of the lattice, respectively. By fitting the data for $B$ = 5 T, $\gamma$, $\beta$, and $\delta$ were estimated as 2.02(2) mJmol$^{-1}$K$^{-2}$, 0.718(4) mJmol$^{-1}$K$^{-4}$, and 0.0204(2) mJmol$^{-1}$K$^{-6}$, respectively. The Debye temperature ($\theta_D$) was calculated from $\beta = (12/5)\pi^4(2N)k_B\theta_D^{-3}$, where $N$ is the Avogadro constant, and $k_B$ is Boltzmann constant; the calculated $\theta_D$ was 175.6(4) K. The electronic contribution of specific heat ($C_{el}$) was calculated by subtracting phonon contribution from the total specific heat for $B$ = 0 T. $C_{el}/T$ was plotted as a function of $T$ in **Fig. 4(b)**. The $C_{el}/T$ shows a clear jump and decreases with decreasing temperature. It suggests the emergence of bulk superconductivity, which is consistent with the result of magnetic susceptibility measurement. $T_c$ and superconducting jump ($\Delta C_{el}$) at $T_c$ were estimated by considering the entropy balance, and the calculated values are $T_c$ = 2.33 K and $\Delta C_{el} = 1.49\gamma T_c$, respectively. This result is comparable to $\Delta C_{el} = 1.43\gamma T_c$ which is expected from the BCS theory with a weak-coupling fully-gapped superconductor [30].

To investigate the carrier type, Seebeck coefficient ($S$) was measured at room temperature for (AgPbBi)$_{(1-x)/3}$In$_x$Te. **Figure 5** shows the nominal In concentration ($x$) dependence of $S$. For $x = 0$, $S$ is -348 μV/K, which suggests that AgPbBiTe$_3$ is a n-type semiconductor. By In substitution, the absolute value becomes very small, even at 10% doping. The $S$ was negative for all compositions, which suggests that electrons are doped by In substitution, and is consistent with the discussion on In valence state in the lattice constant part. Therefore, we conclude that the average In valence is greater than +2 in the present system. This trend in (AgPbBi)$_{(1-x)/3}$In$_x$Te is quite similar to those observed in In-doped AgSnBiTe$_3$ [18]. Since the band structure of (AgPbBi)$_{(1-x)/3}$In$_x$Te expected from the lattice constants possesses band inversion [18], the observation of bulk superconductivity in $(AgPbBi)_{(1-x)/3}In_xTe$ will expand the field to explore exotic superconducting states including topological superconductivity. Therefore, growth of single crystals of $(AgPbBi)_{(1-x)/3}In_xTe$ is needed for the next step, and further experimental and theoretical investigations on $M$Te superconductors are desired.

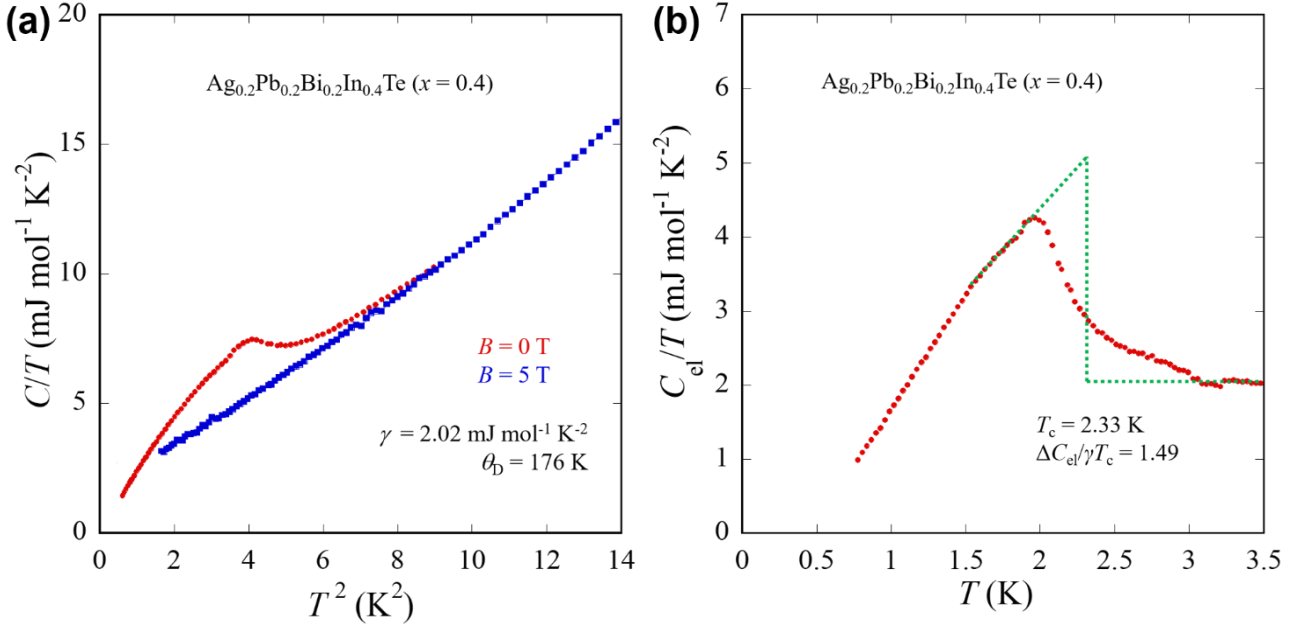

**Fig. 4.** (a) Specific heat ($C/T$) data for $x = 0.4$ plotted as a function of $T^2$ under magnetic fields of 0 and 5 T. (b) Temperature dependence of specific heat of superconducting states ($\Delta C_{el}/T$) for $x = 0.4$. The green dashed line guides the assumed transition by considering the entropy balance.

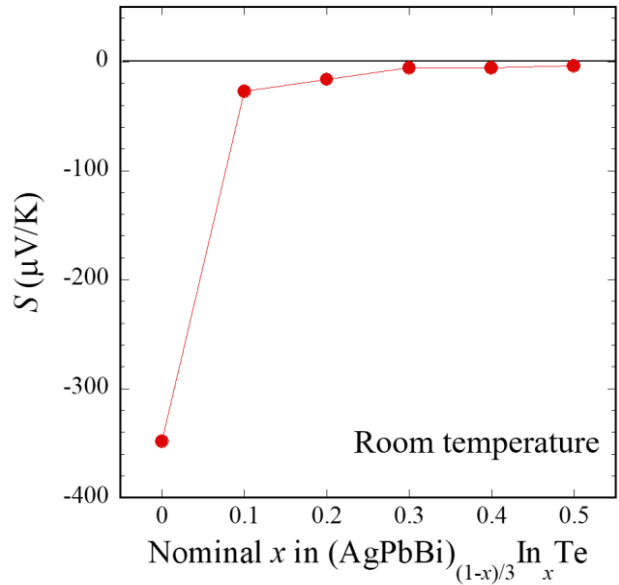

**Fig. 5.** Nominal $x$ dependence of the room-temperature Seebeck coefficient for $(AgPbBi)_{(1-x)/3}In_xTe$.

## 4. Conclusion

We synthesized polycrystalline samples of new NaCl-type $M$Te superconductor (AgPbBi)$_{(1-x)/3}$In$_x$Te ($x$ = 0–0.5). From powder XRD analyses, NaCl-type structure was confirmed for all compositions. The solubility limit of In is $x$ = 0.4–0.5, i.e. 40–50% In for the $M$ site. For $x$ = 0.2–0.5, superconducting transitions were observed in magnetic susceptibility measurements, and the highest $T_c^{mag}$ was 2.8 K for $x$ = 0.4. Each $T_c^{mag}$ is higher than that of (AgSnBi)$_{(1-x)/3}$In$_x$Te system [18], which is consistent with the common trend on the correlation between $T_c$ and lattice constant in $M$Te [18,19,21]. For $x$ = 0.4, the upper critical field was estimated as 1.5 T by GL model fitting of temperature dependence of electrical resistivity under various magnetic field. Bulk nature of the superconductivity was confirmed by the specific heat jump at $T_c$. From the evolution of Seebeck coefficient by In doping and the deviation of $x$ dependence of lattice constant from the Vegard's law, we concluded that the average In valence is greater than +2, and electron carriers are generated in (AgPbBi)$_{(1-x)/3}$In$_x$Te system by In doping.

## CRediT authorship contribution statement

**Takahiro Sawahara:** Data curation, Formal analysis, Investigation, Methodology, Validation, Visualization, Writing - original draft

**Hiroto Arima:** Data curation, Formal analysis, Investigation, Writing - review & editing

**Takayoshi Katase:** Investigation, Resources, Writing - review & editing

**Aichi Yamashita:** Data curation, Formal analysis, Investigation, Methodology, Resources, Supervision, Visualization, Writing - review & editing

**Ryuji Higashinaka:** Investigation, Resources, Writing - review & editing

**Yoshikazu Mizuguchi:** Conceptualization, Data curation, Formal analysis, Funding acquisition, Investigation, Methodology, Project administration, Resources, Supervision, Validation, Visualization, Writing - original draft, Writing - review & editing

## Declaration of Competing Interests

The authors declare no competing interests.

## Acknowledgements

The authors thank A. Seshita, M. R. Kasem, Y. Nakahira, O. Miura, T. D. Matsuda, Y. Aoki, K. Hoshi, D. Gomita, and Y. Goto for their supports in experiments and fruitful discussion. This work was partly supported by JSPS-KAKENHI (Grant No.: 21H00151), Tokyo Government Advanced Research (Grant No.: H31-1), JST-ERATO (JPMJER2201), and the Collaborative Research Project of Laboratory for Materials and Structures, Institute of Innovative Research, Tokyo Institute of Technology.

**Reference**

[1] Y. Tanaka, Zhi Ren, T. Sato, K. Nakayama, S. Souma, T. Takahashi, K. Segawa, Y. Ando, Experimental realization of a topological crystalline insulator in SnTe, Nat. Phys. 8 (2012) 800-803.
https://doi.org/10.1038/nphys2442

[2] T. H. Hsieh, H. Lin, J. Liu, W. Duan, A. Bansil, L. Fu, Topological crystalline insulators in the SnTe material class, Nat. Commun. 3 (2012) 982.
https://doi.org/10.1038/ncomms1969

[3] Y. Ando, Topological Insulator Materials, J. Phys. Soc. Jpn. 82 (2013) 102001.
https://doi.org/10.7566/JPSJ.82.102001

[4] R. J. Cava, H. Ji, M. K. Fuccillo, Q. D. Gibsona, Y. S. Horb, Crystal structure and chemistry of topological insulators, J. Mater. Chem. C 1 (2013) 3176-3189.
https://doi.org/10.1039/C3TC30186A

[5] Y. Xiao, L. D. Zhao, Charge and phonon transport in PbTe-based thermoelectric materials, npj Quantum Mater. 3 (2018) 55.
https://doi.org/10.1038/s41535-018-0127-y

[6] S. Ching-Hua, Design, growth and characterization of PbTe-based thermoelectric materials, Prog. Cryst. Growth Charact. Mater. 65, (2019) 47-94.
https://doi.org/10.1016/j.pcrysgrow.2019.04.001

[7] Y. Gelbstein, Z. Dashevsky, M.P. Dariel, High performance n-type PbTe-based materials for thermoelectric applications, Physica B 363, 1-4 (2015) 196-205.
https://doi.org/10.1016/j.physb.2005.03.022

[8] S. Sportouch, M. Basteat, P. Brazis, J. Ireland, C. R. Kannewurf, C. Uher, M. G. Kanatzidis, Thermoelectric Properties of the Cubic Family of Compounds $AgPbBiQ_3$ (Q = S, Se, Te). Very Low Thermal Conductivity Materials, MRS Online Proc. Lib. 545 (1998) 123-130.
https://doi.org/10.1557/PROC-545-123

[9] R. Knura, T. Parashchuk, A. Yoshiasa, K. T. Wojciechowski, Origins of low lattice thermal conductivity of $Pb_{1-x}Sn_xTe$ alloys for thermoelectric applications, Dalton Trans. 50 (2021) 4323-4334.
https://doi.org/10.1039/D0DT04206D

[10] H. E. BÖMMEL, A. J. DARNELL, W. F. LIBBY, B. R. TITTMANN, A. J. YENCHA, Superconductivity of Metallic Indium Telluride, Science 141, (1963) 714.
https://doi.org/10.1126/science.141.3582.714.a

[11] S. Geller, A. Jayaraman, G. W. Hull, Jr. SUPERCONDUCTIVITY AND VACANCY STRUCTURES OF THE PRESSURE-INDUCED NaCl-TYPE PHASES OF THE In–Te SYSTEM, Appl. Phys. Lett. 4 (1964) 35–37.
https://doi.org/10.1063/1.1753952

[12] P. B. Allen, M. L. Cohen, Carrier-Concentration-Dependent Superconductivity in SnTe and GeTe, Phys. Rev. 177 (1969) 704.
https://doi.org/10.1103/PhysRev.177.704

[13] A. S. Erickson, J.-H. Chu, M. F. Toney, T. H. Geballe, I. R. Fisher, Enhanced superconducting pairing interaction in indium-doped tin telluride, Phys. Rev. B 79 (2009) 024520.
https://doi.org/10.1103/PhysRevB.79.024520

[14] G. Balakrishnan, L. Bawden, S. Cavendish, M. R. Lees, Superconducting properties of the In-substituted topological crystalline insulator SnTe, Phys. Rev. B 87 (2013) 140507(R).
https://doi.org/10.1103/PhysRevB.87.140507

[15] M. Novak, S. Sasaki, M. Kriener, K. Segawa, Y. Ando, Unusual nature of fully gapped superconductivity in In-doped SnTe, Phys. Rev. B 88 (2013) 140502(R).
https://doi.org/10.1103/PhysRevB.88.140502

[16] N. Haldolaarachchige, Q. Gibson, W. Xie, M. B. Nielsen, S. Kushwaha, R. J. Cava, Anomalous composition dependence of the superconductivity in In-doped SnTe, Phys. Rev. B 93 (2016) 024520.
https://doi.org/10.1103/PhysRevB.93.024520

[17] Y. Mizuguchi, O. Miura, High-Pressure Synthesis and Superconductivity of Ag-Doped Topological Crystalline Insulator SnTe ($Sn_{1-x}Ag_xTe$ with x = 0–0.5), J. Phys. Soc. Jpn. 85 (2016) 053702.
https://doi.org/10.7566/JPSJ.85.053702

[18] T. Mitobe, K. Hoshi, M. R. Kasem, R. Kiyama, H. Usui, A. Yamashita, R. Higashinaka, T. D. Matsuda, Y. Aoki, T. Katase, Y. Goto, Y. Mizuguchi, Superconductivity in In-doped AgSnBiTe3 with possible band inversion, Sci. Rep. 11 (2021) 22885.
https://doi.org/10.1038/s41598-021-02341-9

[19] M. Katsuno, R. Jha, K. Hoshi, R. Sogabe, Y. Goto, Y. Mizuguchi, High-Pressure Synthesis and Superconducting Properties of NaCl-Type $In_{1-x}Pb_xTe$ (x = 0–0.8), Condens. Matter 5 (2019) 14.
https://doi.org/10.3390/condmat5010014

[20] M. R. Kasem, Y. Nakahira, H. Yamaoka, R. Matsumoto, A. Yamashita, H. Ishii, N. Hiraoka, Y. Takano, Y. Goto, Y. Mizuguchi, Robustness of superconductivity to external pressure in high-entropy-alloy-type metal telluride $AgInSnPbBiTe_5$, Sci. Rep. 12 (2022) 7789.
https://doi.org/10.1038/s41598-022-11862-w

[21] M. R. Kasem, R. Ishi, T. Katase, O. Miura, Y. Mizuguchi, Tuning of carrier concentration and superconductivity in high-entropy-alloy-type metal telluride $(AgSnPbBi)_{(1-x)/4}In_xTe$, J. Alloys Compd. 920 (2022) 166013.
https://doi.org/10.1016/j.jallcom.2022.166013

[22] A. Yamashita, R. Jha, Y. Goto, T. D. Matsuda, Y. Aoki, Yoshikazu Mizuguchi, An efficient way of increasing the total entropy of mixing in high-entropy-alloy compounds: a case of NaCl-type $(Ag,In,Pb,Bi)Te_{1-x}Se_x$ (x = 0.0, 0.25, 0.5) superconductors, Dalton Trans. 49 (2020) 9118-9122.
https://doi.org/10.1039/D0DT01880E

[23] Y. Mizuguchi, Superconductivity in High-Entropy-Alloy Telluride $AgInSnPbBiTe_5$, J. Phys. Soc. Jpn. 88 (2019) 124708.
https://doi.org/10.7566/JPSJ.88.124708

[24] Y. Matsushita, P. A. Wianecki, A. T. Sommer, T. H. Geballe, I. R. Fisher, Type II superconducting parameters of Tl-doped PbTe determined from heat capacity and electronic transport measurements, Phys. Rev. B 74 (2006) 134512.
https://doi.org/10.1103/PhysRevB.74.134512

[25] G. Tan, F. Shi, H. Sun, L. D. Zhao, C. Uher, V. P. Dravid, M. G. Kanatzidis, SnTe–AgBiTe$_2$ as an efficient thermoelectric material with low thermal conductivity, J. Mater. Chem. A 2 (2014) 20849-20854
https://doi.org/10.1039/C4TA05530F

[26] V. Petříček, M. Dušek, L. Palatinus, Crystallographic Computing System JANA2006: General features, Z. Kristallogr. - Cryst. Mater. 229 (2014) 345–352.
https://doi.org/10.1515/zkri-2014-1737

[7] K. Momma, F. Izumi, VESTA: a three-dimensional visualization system for electronic and structural analysis, J. Appl. Cryst. 41 (2008) 653-658.
https://doi.org/10.1107/S0021889808012016

[28] V. P.S. Awana, R. S. Meena, A. Pal, A. Vajpayee, K. V.R. Rao, H. Kishan, Physical property characterization of single step synthesized NdFeAsO$_{0.80}$F$_{0.20}$ bulk 50 K superconductor, Eur. Phys. J. B. 79 (2011) 139-146.
https://doi.org/10.1140/epjb/e2010-10674-x

[29] R. Sultana, P. Rani, A. K. Hafiz, Reena Goyal, V. P. S. Awana, An Intercomparison of the Upper Critical Fields (H$_{c2}$) of Different Superconductors YBa$_2$Cu$_3$O$_7$, MgB$_2$, NdFeAsO$_{0.8}$F$_{0.2}$, FeSe$_{0.5}$Te$_{0.5}$ and Nb$_2$PdS$_5$, J. Supercond. Novel Magn. 29 (2016) 1399-1404.
https://doi.org/10.1007/s10948-016-3507-1

[30] J. Bardeen, L. N. Cooper, J. R. Schrieffer, Microscopic Theory of Superconductivity, Phys. Rev. 106 (1957) 162.
https://doi.org/10.1103/PhysRev.106.162